\documentclass[%
 reprint,
superscriptaddress,
 amsmath,amssymb,
 aps,
 prl,
]{revtex4-1}

\usepackage[dvipsnames]{xcolor}

\usepackage{graphicx}
\usepackage{dcolumn}
\usepackage{bm}

\begin{document}

\noindent
{\bf O'Connor, Alvarez and Robbins Reply:} 
The preceding Comment erroneously applies the entropic stress expression in our Letter \cite{TOC2018} to transient stress.
In addition,
the authors only apply this expression at extreme extension rates where Ref. [1] clearly showed deviations from the entropic stress expression for steady-state extensional flow.
Hence the surprisingly minor discrepancies noted in the Comment between observed and ``predicted'' stress are entirely expected and have no bearing on the discussion or conclusions in our Letter \cite{TOC2018}.

Our Letter \cite{TOC2018} developed an explanation for puzzling trends in steady-state measurements \cite{Bach2003,Sridhar2013,Wingstrand2015} of the rate-dependent extensional viscosity of entangled polymer melts. Simulations of polymers with different length and entanglement density revealed a crossover between two limiting behaviors.
The linear response was accurately described by tube theory using only previously published entanglement times $\tau_e$ and entanglement lengths $N_e$. A new expression for the high-rate viscosity was derived that collapsed all high-rate simulation data. The linear and high-rate responses scale with different powers of chain length and have different drag coefficients and stiffness dependence. These differences explain trends in the amount of extension-rate thinning for different polymers in our simulations and in experiments \cite{Bach2003,Sridhar2013,Wingstrand2015}.

Reference \cite{TOC2018} also described how the tube confining polymers aligned, stretched and narrowed with elongational rate. Changes in alignment were shown to collapse when rate was normalized by the equilibrium disentanglement time from tube theory, while changes in tube length and radius depended only on the equilibrium Rouse time. Our central result about the scaling of high-rate viscosity relies only on the observation that chains are nearly straight at high rates. The preceding Comment does not make any statement that questions the main results described above.

An independent point was a comparison of the steady-state macroscopic stress
$\sigma_{ex}$ and an entropic stress $\sigma_{ex}^{ent}$
associated with the loss in entropy of stretched
segments with length equal to the equilibrium $N_e$.
The major goal was to demonstrate that the equilibrium $N_e$ remains relevant
even in far-from-equilibrium flow.
Fig. 4 of Ref. \cite{TOC2018} compared the two stresses
using previously published  values of the Kuhn length $\ell_K$ and $N_e$.
Results for all chain lengths and stiffnesses collapsed for stresses from 0.003 to 2, which corresponds to 0.015 to 100MPa using common mappings to real units \cite{Ge14a}.
Over this range, deviations are less than 10\% for one chain stiffness and 30\% for the other. We noted that deviations became ``significant'' (up to 40\%) at the very highest stresses (up to 200MPa) where chains are nearly straight and the analytic expression for entropy in Eqs. 1 of both Ref. [1] and the Comment is becoming inadequate. We did not have space to go into more detail, but Eq. 1 is known to become inaccurate when chains are pulled taut on the Kuhn length scale \cite{Blume, Milner}. This leads to energetic corrections that are entirely consistent with the observations in the Comment and the percentage errors noted above.
Experiments are usually unable to reach this extreme limit. For example, the largest stresses obtained for polystyrene are less than 10MPa. The preceding Comment says ``data in Fig. 4 of Ref. [1] indicate a lack of ``quantitative'' agreement at tensile stress higher than 0.1.'' The authors give no justification for this statement, but we note that experiments are typically below this stress.

The Comment extends Eq. 1 from the steady-state regime considered in Ref. [1], to the transient behavior during startup.
In personal communications we have tried to make clear to the authors of the Comment that we do not believe the entropic stress should be quantitatively accurate during start up and yet they label a curve on their plot with our names.
Given that they only show results for a rate where we noted our steady state errors were significant, the values of macroscopic and entropic stress evolve in strikingly similar ways in their figure.
This is particularly surprising given that they plot stress at intervals corresponding to about the entanglement time and steady state is reached at only 10\% of the Rouse time.
We did not apply the concept of equilibrium entropy to such systems because chain conformations are evolving too rapidly.
The Comment says ``there should be no ambiguity in calculating the classical intrachain entropic stress in the small strain limit.'' That statement is manifestly false when entropic forces are evaluated more quickly than an ensemble of chain conformations can be sampled.

In conclusion, the preceding Comment has little to do with the points made in our Letter. The energetic terms they discuss are known corrections to the force required to stretch a chain segment when the force exceeds $k_B T/\ell_K$ \cite{Blume,Milner}.
These corrections are small for most of the range of experimental interest.
The bulk of their Comment refers to transient effects that are not related to our Letter and ignores important nonequilibrium effects.

\begin{acknowledgments}
This work was performed within the Center for Materials in Extreme Dynamic Environments 
with financial support provided by grant No. W911NF-12-2-0022.
\end{acknowledgments}
%

\end{document}